\documentclass[prb,aps,twocolumn,floatfix]{revtex4}
\usepackage{graphicx,bm}
\usepackage{subfigure}
\usepackage{comment}
\usepackage{amssymb}
\usepackage{verbatim}
\begin{document}
\pagestyle{plain}
\title{Adiabatic Domain Wall Motion and Landau-Lifshitz Damping}
\author{M. D. Stiles$^\ddagger$, W.M. Saslow$^\S$,
M. J. Donahue$^\dagger$, and A. Zangwill$^\P$ \vspace{.5em}}

\affiliation{$^\ddagger$Center for Nanoscale Science and Technology, National
  Institute of Standards and Technology, Gaithersburg, MD 20899-8412 \vspace{.5em} }
\affiliation{$^\S$Department of Physics, Texas A\&M University, College
  Station, TX 77843-4242 \vspace{.5em}}

\affiliation{$^\dagger$Mathematical and Computational Sciences Division, National
  Institute of Standards and Technology, Gaithersburg, MD 20899-8910 \vspace{.5em}}

\affiliation{$^\P$School of Physics, Georgia Institute of
Technology, Atlanta, GA 30332-0430 \vspace{.5em}}
\begin{abstract}
Recent theory and measurements of the velocity of current-driven
domain walls in magnetic nanowires have re-opened the unresolved
question of whether Landau-Lifshitz damping or Gilbert damping
provides the more natural description of dissipative magnetization
dynamics.  In this paper, we argue that (as in the past) experiment
cannot distinguish the two, but that Landau-Lifshitz damping
nevertheless provides the most physically sensible interpretation of
the equation of motion.  From this perspective, (i) adiabatic
spin-transfer torque dominates the dynamics with small corrections
from non-adiabatic effects; (ii) the damping always decreases the
magnetic free energy, and (iii) microscopic calculations of damping
become consistent with general statistical and thermodynamic
considerations.
\end{abstract}
\date{\today}
\maketitle

\section{Background}

Experiments designed to study the effect of electric current on domain
wall motion in magnetic nanowires show that domain walls move over
large distances with a velocity proportional to the applied
current.\cite{Koo:2002,Tsoi:2003,Klaui:2003,Grollier:2003,Vernier:2004,Yamaguchi:2004,Lim:2004,Klaui:2005,Hayashi:2006,Beach:2006}
Most theories ascribe this behavior to the interplay between {\it
spin-transfer} (the quantum mechanical transfer of spin angular
momentum between conduction electrons and the sample magnetization)
and magnetization damping of the Gilbert type.\cite{Gilbert} Contrary
to the second point, we argue in this paper that Landau-Lifshitz
damping\cite{LL} provides the most natural description of the
dynamics. This conclusion is based on the premises that damping should
always reduce magnetic free energy and that microscopic calculations
must be consistent with statistical and thermodynamic considerations.

Theoretical studies of current-induced domain wall motion typically focus 
on one-dimensional models where current flows in the $x$-direction
through a magnetization ${\bf M}(x)=M\hat{\bf M}(x)$. When 
$M$ is constant, the equation of
motion is
\begin{equation}
\label{DSZ1}
{\dot{\bf M}} = -\gamma {\bf M} \times {\bf H} + {\bf N}_{\rm ST} +{\bf D}.
\end{equation}
The precession torque $-\gamma {\bf M} \times {\bf H}$ depends on the
gyromagnetic ratio $\gamma$ and an effective field $\mu_0{\bf H}=-\delta
F/\delta {\bf M}$ which accounts for external fields, anisotropies,
and any other effects that can be modelled by a free energy $F[{\bf
M}]$ ($\mu_0$ is the magnetic constant). The spin-transfer torque
${\bf N}_{\rm ST}$ is not derivable 
from a potential, but its form is fixed by symmetry arguments and
model calculations.\cite{Berger:1978,Bazaliy:1998,Ansermet:2004,Tatara:2004,Waintal:2004,ZL:2004,Thiaville:2005,Dugaev:2005,XZS:2006}
A local approximation \cite{caveat} (for current in the $x$-direction) is
\begin{equation}
\label{DSZ2}
    {\bf N}_{\rm ST} = -\upsilon \left[\partial_x{\bf M}  
- \beta \hat{\bf M}\times\partial_x{\bf M}\right] .
\end{equation}
The first term in (\ref{DSZ2}) occurs when the spin current 
follows the domain wall magnetization adiabatically, {\it i.e.}, when the 
electron spins remain largely aligned (or antialigned) with the 
magnetization as they propagate through the wall.
 The constant $\upsilon$ is a velocity. If $P$ is the spin 
 polarization of the current, $j$ is the current 
density, and $\mu_B$ is the Bohr magneton,   
\begin{equation}
\label{speed}
\upsilon={-Pj\mu_B\over eM}.
\end{equation}
The second term in 
(\ref{DSZ2}) arises 
from non-adiabatic effects. The constant 
$\beta$ is model dependent.

The damping torque ${\bf D}$ in (\ref{DSZ1}) accounts for 
dissipative processes, see \onlinecite{Heinrich} for a review. Two  
phenomenological  
forms for ${\bf D}$ are  
employed commonly: the Landau-Lifshitz form\cite{LL} with damping constant
$\lambda$,
\begin{equation}
\label{DL}
{\bf D}_L= -  \lambda {\bf\hat{ M}} \times 
                    \left( {\bf M} \times {\bf H} \right),
\end{equation}
and the Gilbert form\cite{Gilbert} with damping constant $\alpha$,
\begin{equation}
\label{DG}
{\bf D}_G= \alpha {\bf \hat{M}} \times {\dot{\bf M}}.
\end{equation}
The difference between the two is usually very small and almost all
theoretical and simulation studies of current-induced domain wall
motion solve (\ref{DSZ1}) with the Gilbert form of
damping.\cite{ZL:2004,Thiaville:2005,Li:2004,Thiaville:2004,He:2005,Dugaev:2005,Tatara:2005,He:2006}
This is significant because, as we now discuss, Gilbert damping and
Landau-Lifshitz damping produce quite different results for this
problem when the same spin transfer torque is used.

Consider a  N\'{e}el wall where ${\bf M}$ lies entirely in the
plane of a thin film when the current is zero. By definition,
$\hat{\bf M}\times{\bf H}=0$ if we choose ${\bf M}(x)$ as the
equilibrium structure which minimizes the free energy $F[{\bf M}]$. The wall
distorts if  $\hat{\bf M}\times{\bf H}\neq 0$ for any
reason. The theoretical literature cited above shows that, with
damping omitted, the  N\'{e}el wall moves undistorted at the speed
$\upsilon$  [see (\ref{speed})] when $\beta=0$ in (\ref{DSZ2}).
Gilbert damping brings this motion to a stop because ${\bf D}_G$
rotates ${\bf M}(x)$ out-of-plane until the torque from magnetostatic
shape anisotropy cancels the spin-transfer torque.  However, if the
non-adiabatic term in (\ref{DSZ2}) is non-zero, steady wall motion
occurs at  speed $\beta \upsilon /\alpha$. 

Using this information, two recent
experiments\cite{Hayashi:2006,Beach:2006} 
used their  observations of average domain wall velocities very near
$\upsilon$ to infer that $\beta \approx \alpha$ for permalloy
nanowires. 
This is consistent with microscopic calculations (which include
disorder-induced spin-flip scattering) that report $\beta=\alpha$
\cite{Tserkovnyak:2006} or $\beta\approx\alpha$ \cite{Kohno:2006} for
realistic band models of an itinerant ferromagnet.  On the other hand,
calculations for ``s-d'' models of ferromagnets with localized moments
find little numerical relationship between $\beta$ and $\alpha$
\cite{Kohno:2006,Duine:2007}.

A rather different interpretation of the data follows from a
discussion of current-driven domain wall motion in the s-d model
offered by Barnes and Maekawa.\cite{Barnes:2005} These authors argue
that there should be no damping of the magnetization when a wall which
corresponds to a minimum of the free energy $F[{\bf M}]$
simply translates at constant speed. This is true of ${\bf D}_L$ in
(\ref{DL}) because ${\bf M}\times{\bf H}=0$ but it is {\it not} true
of ${\bf D}_G$ 
in (\ref{DG}) because ${\bf \dot{M}}\neq 0$ when ${\bf N}_{\rm ST}\neq
0$. From this point of view, the ``correct'' equation of motion is
\begin{equation}
\label{correct}
{\dot{\bf M}} = -\gamma {\bf M} \times {\bf H} -\upsilon \partial_x {\bf M}  -  \lambda {\bf\hat{ M}} \times 
                    \left( {\bf M} \times {\bf H} \right),
\end{equation}
because it reduces (for energy-minimizing walls) to  
\begin{equation}
\label{mini}
{\dot {\bf M}}= -\upsilon\partial_x {\bf M}.
\end{equation}
In the absence of extrinsic pinning, this argument identifies the
experimental observation of long-distance wall motion with a uniformly
translating solution ${\bf M}(x-\upsilon t)$ of (\ref{mini}) with
minimum energy.

As we discuss below, it is possible to convert between descriptions
with Landau-Lifshitz and Gilbert dampings by concurrently changing the
value of the non-adiabatic spin-transfer torque.  The Landau-Lifshitz
description in Eq.~\ref{correct} is equivalent to one with Gilbert damping
with $\beta=\alpha$.  The goal of this paper is to argue that there
are conceptual reasons to prefer the description with Landau-Lifshitz
damping even when $\beta\ne\alpha$.

Section II presents micromagnetic simulations that confirm the
discussion above and describes further details. Then,
the remainder of this paper provides three theoretical arguments which
support the use Landau-Lifshitz damping for current-driven domain wall
motion (in particular) and for other magnetization dynamics problems
(in general).  First, we reconcile our preference for Landau-Lifshitz
damping with the explicit microscopic calculations of Gilbert damping
and non-adiabatic spin torque reported in
Refs.~\onlinecite{Tserkovnyak:2006,Kohno:2006,Duine:2007}. Second, we
show that Gilbert damping can increase the magnetic free energy in the
presence of spin-transfer torques.  Finally, we show that
Landau-Lifshitz damping is uniquely selected for magnetization
dynamics when the assumptions of non-equilibrium thermodynamics are
valid.

\section{Micromagnetics}

\begin{figure}
\centerline{\includegraphics[width=0.5\textwidth]{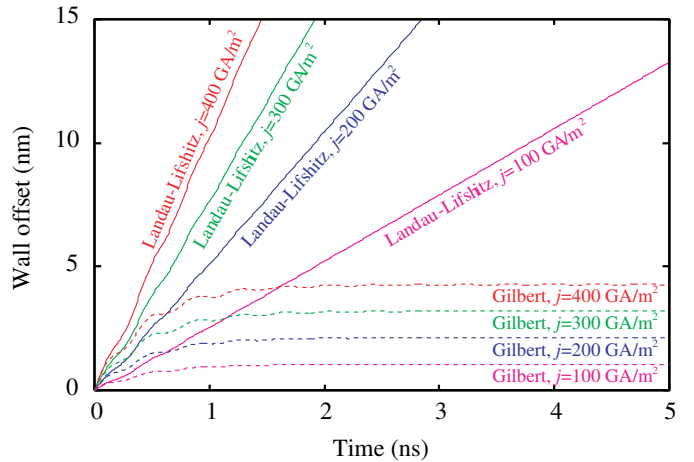}}
\caption{Position versus time for a transverse domain wall and several
  values of the applied current density computed with adiabatic spin
  torques ($\beta=0$) and the two forms of damping in (\ref{DL}) and
  (\ref{DG}). }
\label{transverse}
\end{figure}

Our analysis begins with a check on the robustness of the foregoing
model predictions using full three-dimensional micromagnetic
simulations of current-driven domain wall motion.\cite{OOMMF} We
studied nanowires 12~nm thick and 100~nm wide with material parameters
chosen to simulate ${\rm Ni}_{80}{\rm Fe}_{20}$. At zero current, this
geometry and material system support in-plane magnetization with
stable domain walls of transverse
type.\cite{Donahue:1997} Figure~\ref{transverse} shows the wall
position as a function of time for a transverse domain wall for
several values of applied current density $j$.  The curves labelled
Gilbert ($\alpha=0.02$) show that wall motion comes quickly to a
halt. Examination of the magnetization patterns confirms the torque
cancellation mechanism outlined above.  The curves labelled
Landau-Lifshitz show that the wall moves uniformly with the velocity
given by 
(\ref{speed}) which is independent of the damping parameter
$\lambda$.\cite{caveat2}

The sudden turn-on of the current and hence Oersted magnetic field at
$t=0$ generates the small amplitude undulations of the curves in
Figure~\ref{transverse} but otherwise has little effect on the
dynamics. 
An initial state of a stable vortex wall in a 300 nm wide wire
produces similar results, except that under the Gilbert formulation
the vortex wall moves about twenty times farther before stopping as
compared to the transverse wall in the 100 nm wire.
We conclude from these simulations that the basic picture of
domain wall dynamics gleaned from one-dimensional models is correct.

The magnetic free energy behaves differently in simulations depending
on whether Landau-Lifshitz or Gilbert damping is used.  Before the
current is turned on, the domain wall is in a configuration that is a
local minimum in the energy.  For Landau-Lifshitz damping, the energy
remains largely constant near this minimum and is exactly constant if
the Oersted fields are ignored.  For Gilbert damping, the energy
increases when the current is turned on and the walls distort.  For a
transverse wall, the distortion is largely an out of plane tilting.
Initially, the energy increases at a rate proportional to the damping
parameter (ignoring higher order corrections discussed in the next
section).  The details of this behavior are somewhat obscured by the
oscillations due to the Oersted magnetic field, but are quite apparent
in simulations in which this field is omitted.  As the wall tilts out
of plane, the torque due to the magnetostatic field opposes the wall
motion and the wall slows down.  Eventually the torque balances the
adiabatic spin transfer torque and the wall stops.

In simulations using Gilbert damping, the change in magnetic free
energy between the initial and final configurations is independent of
the damping parameter as it is determined by the balance between the
magnetostatic torque and the adiabatic spin transfer torque.  However,
the amount of time before the wall stops and the distance the wall
moves are inversely proportional to the damping parameter.  The
Gilbert damping torque is responsible for this increase in energy as
can be seen from analyzing the directions of the other torques.
Precessional torques, like those due to the exchange and the
magnetostatic interactions that are important in these simulations, by
their nature are directed in constant energy directions and do not
change the magnetic free energy.  The adiabatic spin transfer torque
is in a direction that translates the domain wall and does not change
the energy in systems where the energy does not depend on the position
of the wall.  Thus, in simulations of ideal domain wall motion without
Oersted fields, the Gilbert damping torque is the only torque that
changes the energy.  Throughout these simulations, the Gilbert damping
torque is in a direction that increases rather than decreases the
magnetic free energy.

\section {Magnetic Damping With Spin-Transfer Torque} 

When ${\bf N}_{\rm ST}=0$, it is well known that 
a few lines of algebra converts the
equation of motion (\ref{DSZ1}) with Gilbert damping into
(\ref{DSZ1}) with Landau-Lifshitz damping (and vice-versa) with
suitable redefinitions of the precession constant $\gamma$ and the
damping constants $\lambda$ and $\alpha$.\cite{Bertotti} 
The same algebraic manipulations \cite{Tserkovnyak:2006} show that
(\ref{correct}) is 
mathematically equivalent to a Gilbert-type equation with
$\alpha=\lambda/\gamma$:
\begin{equation}
\label{equivG}
\begin{array}{l}
{\dot {\bf M}}=-\gamma(1+\alpha^2) {\bf M}\times {\bf H}+\alpha{\bf \hat{M}}\times \dot{\bf M} \nonumber \\ \\ \hspace{.35in}
 -\upsilon \left [\partial_x {\bf M}-\alpha {\bf \hat{M}}\times \partial_x {\bf M}\right]. \\ \end{array}  
\end{equation}
To analyze (\ref{equivG}), we first ignore spin-transfer (put
$\upsilon=0$) and note that this re-written Landau-Lifshitz equation
differs from the conventional Gilbert equation only by an
$O(\alpha^2)$ renormalization of the gyromagnetic ratio.
Consequently, first-principles derivations of any equation of motion
for the magnetization must be carried to second order in the putative
damping parameter if one hopes to distinguish Landau-Lifshitz damping
from Gilbert damping. 
This observation shows that papers 
that derive
Gilbert
damping\cite{Kambersky:1970,pump,Tserkovnyak:2006,Kohno:2006,Duine:2007,Koopmans:2005}
or Landau-Lifshitz damping\cite{Callen:1958,Fredkin:2000} from
microscopic calculations carried out only to first order in
$\alpha$ cannot be used to justify one form of damping over the other. 

Now restore the spin-transfer terms in (\ref{equivG}) and note that
the transformation to this equation from (\ref{correct}) automatically
generates a non-adiabatic torque with $\beta=\alpha$.  This
transformation means that to lowest order in $\alpha$ and $\beta$, an
equation of motion with Gilbert damping and a non-adiabatic
coefficient $\beta_{\rm G}$ is equivalent to an equation of motion
with Landau-Lifshitz damping with non-adiabatic coefficient
$\beta_{\rm LL}= \beta_{\rm G}-\alpha$.  This shows that equivalent
equations of motion can be made using either form of damping, albeit
with rather different descriptions of current induced domain wall
motion.  Nevertheless, as we argue below, there are conceptual
advantages to the Landau-Lifshitz form.

\section{Landau-Lifshitz Damping Uniquely Reduces Magnetic Free
  Energy}

Landau-Lifshitz damping irreversibly reduces magnetic free energy when
spin-transfer torque is present. The same statement is not true for
Gilbert damping. This can be seen from the situation described in
Section~II where Gilbert damping causes a minimum energy domain wall
configuration to distort and tilt out of plane. Nothing prevents an
increase in magnetic free energy for this open system, but it is
clearly preferable if changes magnetic configurations that increase
$F[{\bf M}]$ come from the effects of spin-transfer torque rather than
from the effects of a torque intended to model dissipative processes.
This is an important reason to prefer ${\bf D}_L$ in (\ref{DL}) to
${\bf D}_G$ in (\ref{DG}).
This argument depends crucially on the fact that the
adiabatic spin-transfer torque is {\it not} derivable from a free
energy as we discuss henceforth. 

The field ${\bf H}$ in Eq.~(\ref{DSZ1}) is the (negative) gradient of
the magnetic free energy.  The component of this gradient in the
direction that does not change the size of the magnetization is $-
\hat{\bf M}\times[{\bf M}\times{\bf H}]$.  Since this direction is
exactly that of the Landau-Lifshitz form of the damping,
Eq.~(\ref{DL}), it follows that this form of the damping always reduces
this magnetic free energy.  When the Gilbert form of the damping,
Eq.~(\ref{DG}), is used in Eq.~(\ref{DSZ1}), it is possible to rewrite
the damping term as $D_{\rm G} = -\alpha\gamma \hat{\bf M}\times[{\bf M}\times{\bf
H}-(1/\gamma){\bf N}_{\rm ST}]+{\cal O}(\alpha^2)$.  Further, 
one can always write ${\bf
N}_{\rm ST}= -\gamma{\bf M}\times{\bf H}_{\rm ST}$ where ${\bf H}_{\rm
ST}$ is an effective ``spin transfer magnetic field''. However, unlike
the field $\mu_0 {\bf H}=-\delta {\bf F}/\delta {\bf M}$ in (\ref{DSZ1}),
there is no ``spin transfer free energy'' $F_{\rm ST}$ which gives
${\bf H}_{\rm ST}$ as its gradient:
\begin{equation}
\label{nfe}
\mu_0 {\bf H}_{\rm ST}  =  -{\delta F_{\rm ST}\over\delta {\bf M}}~~~~~~~~~~({\rm not~correct}).
\end{equation}
If (\ref{nfe}) were true, the lowest order (in $\alpha$) 
Gilbert damping term $-\alpha\gamma \hat{\bf M}\times[{\bf M}\times({\bf
H}+{\bf H}_{\rm ST})]$ would indeed always lower
the sum $F+F_{\rm ST}$. 
Unfortunately, a clear and convincing demonstration of the
non-conservative nature of the spin-transfer torque is not easy to
find. Therefore, in what follows, we focus on the adiabatic
contribution to (\ref{DSZ2}) and show  that a contradiction arises if
(\ref{nfe}) and its equivalent,     
\begin{equation}
\label{fake}
dF_{\rm ST}=-\mu_0 {\bf H}_{\rm ST}\cdot d{\bf M},
\end{equation}
are true.
\begin{figure}
\centerline{\includegraphics[width=0.5\textwidth]{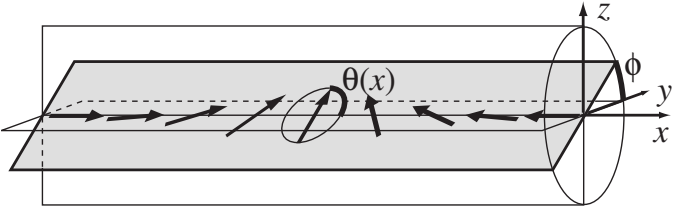}}
\caption{A one-dimensional N\'{e}el domain wall with magnetization ${\bf M}(x)$.}
\label{DomainWall}
\end{figure}

For this argument we consider a simpler model than that discussed in
Section~II. 
Figure~\ref{DomainWall} shows the magnetization ${\bf M}(x)$ for a one
dimensional N\'{e}el wall in a system with uniaxial anisotropy along the
$x$-direction.  The domain wall of width $w$ is centered at $x=0$ and
the plane of the magnetization is tilted out of the $x$-$y$ plane by
an angle $\phi$. A convenient parameterization of the in-plane
rotation angle $\theta(x)$ is  
\begin{equation}
\label{wall}
\theta(x)=\pi/2+\sin^{-1}[\tanh(x/w)].
\end{equation}
Therefore,  
\begin{equation}
\label{Mlabel}
{\bf M}=M[\cos\theta(x),\sin\theta(x)\cos\phi,\sin\theta(x)\sin\phi],
\end{equation}
where $\cos\theta(x) = -\tanh(x/w)$
and 
\begin{equation}
\label{sine}
\sin\theta(x) = {\rm sech}(x/w).
\end{equation}
The magnetic free energy of this domain wall is independent of both its
position and its orientation (angle $\phi$).

For electron flow in the $x$-direction, (\ref{DSZ2}) shows that the 
adiabatic piece of the spin-transfer torque lies entirely in the plane
of the magnetization: 
\begin{equation}
\label{nad}
{\bf N}^{\rm ad}_{\rm ST}\propto \theta'(x)(-\sin\theta,\cos\theta\cos\phi,\cos\theta\sin\phi).
\end{equation}
This torque rotates the magnetization in a manner which produces
uniform translation of the wall in the 
$x$-direction with no change in $\phi$. Since 
\begin{equation}
\label{dt}
\theta'(x)= (1/w) {\rm sech}(x/w),
\end{equation}
comparison with (\ref{sine}) shows that ${\bf N}^{\rm ad}_{\rm ST}=0$
outside the wall as expected.  The magnetic free energy of the domain
wall does not change as the wall is translated.

Now, as indicated above (\ref{nfe}), we are free to interpret the
foregoing wall translation as resulting from local precession of ${\bf
  M}(x)$ around an effective field ${\bf H}_{\rm ST}(x)$ directed
perpendicular to the plane of the domain wall. Specifically,   
\begin{equation}
\label{perp}
{\bf H}_{\rm ST}(x) \propto \theta'(x)(0,-\sin\phi,\cos\phi).
\end{equation}
However, if (\ref{nfe}) and thus (\ref{fake}) are assumed to be
correct, the magnitude and direction of ${\bf H}_{\rm ST}$ imply that
the putative free energy $F_{\rm ST}$ decreases when ${\bf M}(x)$
rotates rigidly around the $x$-axis in the direction of increasing
$\phi$.\cite{porridge} On the other hand, the free energy must return
to its original value when $\phi$ rotates through $2\pi$.  Since the
gradient (\ref{nfe}) can never increase the free energy, we are forced
to conclude that our assumption that $F_{\rm ST}$ exists is incorrect.

\section{A Langevin Equation for the Magnetization}
Neglected work by Iwata\cite{Iwata} treats magnetization dynamics from
the point of view of the thermodynamics of irreversible
processes.\cite{Prig} His non-perturbative calculations uniquely
generates the Landau-Lifshitz form of damping. In this section, we
make equivalent assumptions but go farther and derive an expression
for the damping constant.  Mori and co-workers did this using a
projection operator method.\cite{Mori}  Our more accessible
discussion follows Reif's derivation of a Langevin equation for
Brownian motion.\cite{Reif}

We begin by taking the energy change in a unit volume 
\begin{equation}
dE=- \mu_0 H_{\alpha}dM_{\alpha},
\label{thermo}
\end{equation}
where the repeated index $\alpha$ implies a sum over Cartesian
coordinates. 
It is crucial to note that the magnitude $\vert {\bf M}\vert = M$ is fixed so
only rotations of ${\bf M}$ toward the effective field ${\bf H}$
change the energy of the system.  The interaction with the environment
enters the equation of motion for the magnetization through a
fluctuating torque $N'_{\alpha}$:
\begin{equation}
\frac{dM_{\alpha}}{dt}=-\gamma({\bf M}\times{\bf H})_{\alpha}+N'_{\alpha}.
\label{eqmot}
\end{equation}
The torque ${\bf N}'$ is 
perpendicular to ${\bf M}$ since $|{\bf M}|=M$. 

We consider the evolution of the magnetization over a time interval
$\Delta t$ which is much less than the precession period, but much
greater than the characteristic time scale for the fluctuations
$\tau^{*}$. After this time interval, the statistical average of the
change in magnetization $\Delta M_{\alpha}=M_{\alpha}(t+\Delta
t)-M_{\alpha}(t)$ is
\begin{equation}
\Delta M_{\alpha}=-\gamma({\bf M}\times{\bf H})_{\alpha}(\Delta
t)+\int_t^{t+\Delta t}dt' <N'_{\alpha}(t')>.
\label{Delta_m}
\end{equation}
The equilibrium Boltzmann weighting factor $W_0$ gives 
$<~N'_{\alpha}(t')>_0=0$. 
However, $<~N'_{\alpha}(t')>\ne0$~when~the magnetization is
out of equilibrium. Indeed, this method derives the damping term
precisely from the bias built into the fluctuations due to the changes
$\Delta E=  -\mu_0 H_\nu \Delta M_\nu$ in the energy of the magnetic system.

The Boltzmann weight used to calculate $<N'_{\alpha}(t')>$ is
$W=W_0\exp(- \Delta E/(k_{\rm B}T))$ where (assuming that ${\bf H}$ does not
change much over the integration interval),
\begin{eqnarray}
\Delta E(t')
&=&-\mu_0 H_\nu(t')\int_{t}^{t'}\frac{dM_\nu(t'')}{dt''}dt''
\nonumber\\
&\approx&
-\mu_0 H_\nu(t)\int_{t}^{t'}N'_\nu(t'')dt''.
\label{Delta_E}
\end{eqnarray}
Note that precession 
does not contribute to $\Delta E(t')$. Only motions of
the magnetization that change the energy of the magnetic subsystem
produce bias in the torque fluctuations.  Therefore, since $W=W_0(1-
\Delta E/(k_{\rm B}T))$ for small $\Delta E/(k_{\rm B}T)$, the last
term in (\ref{Delta_m}) now involves only an average over the
equilibrium ensemble:
\begin{eqnarray}
\label{almost}
\Delta M_{\alpha}&\approx&-\gamma({\bf M}\times{\bf H})_{\alpha}(\Delta
t) \\ &+&\frac{\mu_0 H_{\nu}(t)}{k_{\rm B}T} \int_{t}^{t+\Delta t}dt'
\int_{t}^{t'}dt''<N'_{\alpha}(t')N'_{\nu}(t'')>_{0}.\nonumber
\end{eqnarray}

We recall now that the torque fluctuations are correlated over a
microscopic time $\tau^{*}$ that is much shorter than the small but
macroscopic time-interval over which we integrate. Therefore, to the
extent that memory effects are negligible, we define the damping
constant $\lambda$ (a type of fluctuation-dissipation result) from
\begin{equation}
\int_{t}^{t'}dt''<N'_{\alpha}(t')N'_{\nu}(t'')>~\approx~ 
\lambda (k_{\rm B}T M/\mu_0)
\delta^\perp_{\alpha\nu},
\label{correlateT}
\end{equation}
for $|t'- t|\ge\tau^{*}$ and with
$\delta^\perp_{\alpha\nu}= \delta_{\alpha\nu}-\hat{M}_\alpha
\hat{M}_\nu$, which restricts the fluctuations to be transverse to the 
magnetization, but otherwise uncorrelated. 
This approximation reduces the last term in (\ref{almost}) to
$\lambda MH_{\perp\alpha} \Delta t$, where ${\bf H}_\perp=-\hat{\bf
  M}\times(\hat{\bf M}\times {\bf H})$,  is the
piece of ${\bf 
H}$ which is perpendicular to ${\bf M}$. 
Substituting
(\ref{correlateT}) into (\ref{almost}) gives the final result in the form,
\begin{equation}
\label{final}
\frac{d{\bf M}}{dt}\approx-\gamma({\bf M}\times{\bf H})-\lambda
\hat{\bf M}\times({\bf M}\times{\bf H}).
\end{equation}
Equation~(\ref{final}) is the  Landau-Lifshitz equation for the
statistically averaged magnetization. It becomes a Langevin equation when we add a (now) unbiased random torque to the right hand side. 

The procedure outlined above generates higher order terms in $\lambda$
from the expansion of the thermal weighting to higher order in $\Delta
E$.  The second order terms involve an equilibrium average of three powers of
$N'$. These are zero for Gaussian fluctuations.   
The third order terms involve an average of
four powers of $N'$, and are non-zero.  They lead to a term
proportional to $\lambda^{2}H^{2}_{\perp}{\bf H}_{\perp}$, which
we expect to be small and to modify only large-angle
motions of the magnetization. 
\section{Summary}
In this paper, we analyzed current-driven domain wall motion using
both Gilbert-type and Landau-Lifshitz-type damping of the
magnetization motion. Equivalent equations of motion can be written
with either type of damping, but the implied description of the
dynamics (and the relative importance of adiabatic and non-adiabatic
effects) is very different in the two cases.
  
With Landau-Lifshitz damping assumed, adiabatic spin transfer torque
dominates and produces uniform translation of the wall. Non-adiabatic
contributions to the spin transfer torque distort the wall, raise its
magnetic energy, and thus produce a magnetostatic torque which
perturbs the wall velocity. Damping always acts to reduce the
distortion back towards the original minimum-energy wall
configuration.  With Gilbert damping assumed, the damping torque
itself distorts and thereby raises the magnetic energy of the moving
wall. The distortion-induced magnetostatic torque stops domain wall
motion altogether. Additional wall distortions produced by
non-adiabatic spin-transfer torque are needed to produce wall motion.

In our view, Landau-Lifshitz damping is always preferable to Gilbert
damping. When spin-transfer torque is present, this form of damping
inexorably moves the magnetic free energy toward a local
minimum. Gilbert damping does not. Even in the absence of
spin-transfer torque, arguments based on irreversible thermodynamics
show that the Landau-Lifshitz form of damping is uniquely selected for
a macroscopic description.\cite{Iwata} Here, we proceeded equivalently
and derived the Landau-Lifshitz equation of motion as the unique
Langevin equation for the statistical average of a fluctuating
magnetization with fixed spin length.

A.Z. and W.M.S. gratefully acknowledge support from the
U.S. Department of Energy under contracts DE-FG02-04ER46170 and
DE-FG02-06ER46278. 
We thank 
R. A. Duine,
H. Kohno, 
R. D. McMichael, 
J. Sinova, 
N. Smith, 
G. Tatara, and
Y. Tserkovnyak 
for useful discussions.


\end{document}